\documentclass[usegraphicx,usenatbib,useAMS]{mn2e}
\newcommand{\be}{\begin{equation}}
\newcommand{\ba}{\begin{eqnarray}}
\newcommand{\ee}{\end{equation}}
\newcommand{\ea}{\end{eqnarray}}

\def\simless{\mathbin{\lower 3pt\hbox
   {$\rlap{\raise 5pt\hbox{$\char'074$}}\mathchar"7218$}}}   
\def\simgreat{\mathbin{\lower 3pt\hbox
   {$\rlap{\raise 5pt\hbox{$\char'076$}}\mathchar"7218$}}}   
\title[Photoevaporation of cosmological minihaloes II]
{Minihalo photoevaporation during cosmic reionization: evaporation times and
  photon consumption rates}
\author[I.~T.~Iliev et al.]{Ilian~T.~Iliev$^1$, Paul~R.~Shapiro,$^2$ 
and Alejandro~C.~Raga$^3$\\
$^1$ Canadian Institute for Theoretical Astrophysics, University of Toronto, 
60 St. George Street, Toronto, ON M5S 3H8, Canada\\
$^2$ Department of Astronomy, University of Texas, Austin, TX 78712-1083\\
$^3$ Instituto de Ciencias Nucleares,
        Universidad Nacional Autonoma de M\'exico (UNAM),
        Apdo. Postal 70-543, 04510 M\'exico,\\ D. F., M\'exico}

\begin{document}

\maketitle

\label{firstpage}

\begin{abstract}
  The weak, R-type ionization fronts (I-fronts) which swept across the
  intergalactic medium (IGM) during the reionization of the universe often
  found their paths blocked by cosmological minihaloes (haloes with virial
  temperatures $T_{\rm vir}\leq10^4$ K).  When this happened, the neutral gas
  which filled each minihalo was photoevaporated; as the I-front burned its
  way through the halo, decelerating from R-type to D-type, all the halo gas
  was eventually blown back into the IGM as an ionized, supersonic wind.  In a
  previous paper (Shapiro, Iliev \& Raga 2004, hereafter Paper I), we
  described this process and presented our results of the first simulations of
  it by numerical gas dynamics with radiation transport in detail.  
  For illustration we focused on the particular case of a
  $10^7M_\odot$ minihalo which is overrun at $z=9$ by an intergalactic I-front
  caused by a distant source of ionizing radiation, for different types of
  source spectra (either stellar from massive Pop. II or III stars, or 
  QSO-like) and a flux level typical of that expected during reionization. 
  In a Cold Dark Matter (CDM) universe, minihaloes
  formed in abundance before and during
  reionization and, thus, their photoevaporation is an important, possibly
  dominant, feature of reionization, which slowed it down and cost it many
  ionizing photons. In view of the importance of minihalo
  photoevaporation, both as a feedback mechanism on the minihaloes and as an
  effect on cosmic reionization, we have now performed a larger set of
  high-resolution simulations to determine and quantify the dependence of
  minihalo photoevaporation times and photon consumption rates on halo mass,
  redshift, ionizing flux level and spectrum.  We use these results to derive
  simple expressions for the dependence of the evaporation time and photon
  consumption rate on these halo and external flux parameters which can be
  conveniently applied to estimate the effects of minihaloes on the global
  reionization process in both semi-analytical calculations and larger-scale,
  lower-resolution numerical simulations which cannot adequately resolve the
  minihaloes and their photoevaporation. We find that the average number of
  ionizing photons each minihalo atom absorbs during its photoevaporation is
  typically in the range 2-10. For the collapsed fraction in minihaloes
  expected during reionization, this can add $\approx 1$ photon per total atom
  to the requirements for completing reionization, potentially doubling the
  minimum number of photons required to reionize the universe.
\end{abstract}

\begin{keywords}
  hydrodynamics---radiative transfer---galaxies: halos---galaxies:
  high-redshift---intergalactic medium---cosmology: theory
\end{keywords}

\section{Introduction}

The Cold Dark Matter model, due to its hierarchical nature, predicts that a
significant fraction of the matter in the universe before and during
reionization (up to $\sim 30-40\%$ at redshift $z=6$ for current
$\Lambda$CDM)
resided in collapsed and virialized minihaloes - small haloes with virial
temperatures $T_{\rm vir}$ below $10^4$ K. These minihaloes, in the mass range
between the Jeans mass (below which gas pressure forces prevented baryonic
matter in the IGM from collapsing into dark matter haloes as they formed; i.e.
$M_J\simless 10^4M_\odot$) and the mass for which $T_{\rm vir}=10^4$ K
[$M(10^4\, K)\simless10^8M_\odot$] were filled with neutral gas, since their
temperatures were too low for collisional ionization to be effective. Cosmic
reionization had a profound effect on the gas content of these minihaloes and
they, in turn, interfered with cosmic reionization.

This reionization began when the first sources of ionizing radiation turned
on, causing global I-fronts to sweep outward through the surrounding IGM,
creating intergalactic H~II regions, and ended when these H~II regions became
large enough and numerous enough to overlap \citep{SG87}. These global
I-fronts were generally weak, R-type fronts, which moved supersonically, not
only with respect to the cold, neutral gas ahead of them but also with respect
to the gas photoheated to $T\simgreat10^4$K behind them. As such, they raced
ahead of the hydrodynamical response of the ionized gas, unaffected by the
disturbance they created in the gas behind them. That situation changed,
however, when the path of one of these global I-fronts was blocked by a
minihalo. Since the speed of an I-front varies inversely with the density of
the neutral gas ahead of it, the I-front must have slowed down when it
encountered the dense, centrally-concentrated gas inside a minihalo. The
latter is, on average, more than a hundred times denser than the mean IGM,
while at its center it is yet another hundred times denser even than that.
This denser gas slowed the I-front down enough to ``trap'' it, long enough for
the gas-dynamical back-reaction of the disturbed minihalo gas to catch-up to
the I-front and affect its progress. Thereafter, the fate of this I-front and
that of the minihalo gas were inextricably coupled, as ionized minihalo gas,
heated to $T\geq10^4$K, above the binding energy of minihalo gravity, blew
back toward the ionizing source, into the IGM as a supersonic, evaporative
wind.

The first realistic discussion of the photoevaporation of cosmological
minihaloes by global I-fronts during reionization, including the first
radiation-hydrodynamical simulations of this process, was by
\citet{SRM97,SRM98}, developed further in \citet{SR00a,SR00b,SR01,S01} and
Paper I (see Paper I for a more complete description and additional
references.)  As described in Paper I, the dominant sources of reionizing
radiation are likely to have been haloes of virial temperature $T_{\rm
  vir}\geq10^4$K.  Star formation in minihaloes, instead, is thought to have
been inhibited early on by the photodissociating effect on minihalo $H_2$
molecules of the rising UV background just below the H Lyman continuum edge,
created by a small fraction of the sources ultimately required to complete
reionization.  If so, then each source of reionization would typically have
found its sky covered by minihaloes between it and the nearest neighboring
source haloes, and the expansion of I-fronts in the IGM around each source to
the point of overlap would have depended upon the photoevaporation of
intervening minihaloes \citep{HAM01,S01,BL02,SIRM03,SIR04}.  Since the gas
inside collapsed minihaloes was much denser than average, the rate of
recombination there was much higher than average, as well.  Hence, minihalo
photoevaporation was likely to have consumed multiple ionizing photons per
atom, thereby increasing, and potentially dominating, the global photon
consumption during reionization \citep{HAM01,S01,SIR04}.  Therefore,
quantifying this process is crucial for determining the onset, progress and
duration of the reionization of the universe.

An early start ($z \simgreat 15$) and late finish ($z \approx 6$) for cosmic
reionization are suggested by recent observations of the fluctuating
polarization of the CMB ( Kogut et al. 2003) and the Gunn-Peterson effect in
the spectra of quasars at $z \approx 6$ (Becker et al. 2001, Fan et al. 2003),
respectively.  This is consistent with the view that the volume-filling factor
of ionized regions of the universe grew over time from $z \simgreat 15$ until
they overlapped finally at $z \approx 6$.  The presence of minihaloes filled
with neutral gas was affected by this reionization, not only because
photoevaporation stripped pre-existing minihaloes of their baryons, but also
because the subsequent collapse of baryons into newly forming minihaloes was
suppressed by ``Jeans-mass filtering'' of the linear density fluctuations in
the baryonic component after the IGM was heated to $T \simgreat 10^4K$ when it
was reionized (Shapiro, Giroux, and Babul 1994; Shapiro 1995; Gnedin 2000b).
These two processes acted to reduce the presence of neutral-gas-filled
minihaloes only in the ionized regions, however.  The neutral volume outside
these regions continued to form new minihaloes without interruption, at the
unfiltered rate of the universe without reionization.  As a result, the
I-fronts which led the expansion of the ionized regions would have advanced
into neutral gas in which minihaloes formed with ever-increasing abundance
over time.  The unfiltered collapsed fraction in minihaloes for $\Lambda$CDM
at $z = 20$, 15, 9, and 6 was 3, 10, 28, and 36 percent, respectively (Paper
I).  The question of whether minihalo photoevaporation was a dominant feature
of reionization will ultimately be answered only when we know the full story
of cosmic reionization, since there also exist (somewhat more speculative)
scenarios in which the formation of some minihaloes might have been suppressed
by, for example, a significant X-ray background at high-z (e.g. Machacek,
Bryan \& Abel 2003, Glover \& Brand 2003, Madau et al. 2004, Ricotti \&
Ostriker 2004),
double reionization \citep{C03}, entropy floor \citep{OH03}, or reduced
small-scale power in the power spectrum of primordial density fluctuations
\citep{SBL03}.

In Paper I, we discussed the process of photoevaporation of cosmological
minihaloes in detail. We demonstrated the phenomenon of I-front trapping,
whereby the highly supersonic, weak, R-type front propagating in the
low-density IGM slows down upon entering the minihalo, where the gas density
is much higher than that of the IGM, and converts to D-critical and then
subcritical D-type, preceded by a shock which compresses the gas ahead of the
I-front. The I-front slowly propagates subsonically through the halo,
stripping it of gas, layer by layer, as this gas, once ionized,
expands into the IGM as a
supersonic wind towards the ionizing source. This process eventually results
in a dark matter halo completely devoid of gas. We described there in detail
the structure and evolution of the photoevaporative flows and their dependence
on the radiation spectrum of the external ionizing source, considering both
massive stars (Pop II and Pop III) and QSO-like spectra. We derived the
evolution of the position and speed of the I-front, the evolution of the
halo's remaining neutral mass fraction and effective opaque geometric 
cross-section for the absorption of ionizing photons, as well as the 
global parameters of photoevaporation like the 
photoevaporation time, $t_{\rm ev}$, and the number of ionizing photons
absorbed per atom in the course of that evaporation, $\xi$, and its
evolution. Finally, we also discussed some observational diagnostics of the
photoevaporation process. 

In this paper, we extend the results of Paper I by investigating the dependence
of the photoevaporation process on the mass of the halo, on the level and
spectrum of the external photoionizing flux responsible for the I-front which
overtakes the halo, and on the redshift at which this I-front encounters
the minihalo. As discussed in Paper I, we make the assumption that the $\rm
H_2$ molecule formation and cooling which might have led to star formation
inside minihaloes were inefficient in the bulk of the
minihalo population, so the minihaloes were largely ``sterile'' reservoirs of
neutral atomic gas. We concentrate particularly on the dependence of the two
properties of minihalo photoevaporation which are most important for
quantifying their global effect on cosmic reionization, namely, the evaporation
time $t_{\rm ev}$ and the net ionizing photon consumption rate per minihalo
(i.e. the number of ionizing photons absorbed per minihalo atom during the 
evaporation time), $\xi$. In \S~\ref{tev_xi_sect}, we shall define the 
quantities 
$t_{\rm ev}$ and $\xi$ and how we use our simulations to evaluate them more
precisely, and recall some approximations which have been
used previously to estimate these quantities, for comparison.  Our numerical
simulations are described briefly in \S~\ref{calc_sect}. In \S~\ref{tev_sect}
and \S~\ref{xi_sect}, we summarize our numerical results for $t_{\rm ev}$ and 
$\xi$, respectively, for haloes of different masses encountered at different
redshifts by the I-fronts driven by external ionizing sources of different
flux levels and spectra, and compare these with the  
previous estimates described in \S~\ref{tev_xi_sect}. In \S~\ref{Mn_sect}, 
we show that the time-dependence of the evolving neutral mass fraction inside 
a minihalo during its evaporation has a universal shape, independent of halo
parameters and external flux level. The implications of these results for the
theory of cosmic reionization are briefly discussed in
\S~\ref{conclusions_sect}.  Throughout this paper we use the current
concordance model for the background universe and the dark matter - flat,
COBE-normalized $\Lambda$CDM with $\Omega_0=1-\lambda_0=0.3$, $h=0.7$ and 
$\Omega_bh^2=0.02$, with primordial density fluctuation power spectrum
$P(k)\propto k^{n_p}$, with $n_p=1$, and $\sigma_{8h^{-1}}=0.87$
 (e.g. Spergel et al.~2003).

\section{Halo Evaporation Times and Ionizing Photon Consumption Rates}
\label{tev_xi_sect}

We define the evaporation time $t_{\rm ev}$ to be the time at which only $0.1$
per cent of the gas which was initially inside the minihalo remains
neutral\footnote{As discussed in Paper I, the evaporation time is not
  sensitive to the precise value adopted 
  for this remaining neutral fraction in the definition, as long as the latter
  is much less than unity.}.  As in Paper I, we then obtain the number of
ionizing photons absorbed per minihalo atom during this evaporation time by
directly counting the number of recombinations experienced by each of those
initial halo atoms and its first ionization: 
\be 
\xi(t)=\frac{N_{\rm ion}}{N_a}+\frac{1}{N_a}\int_0^{t}dt \int
dV(\alpha_{\rm H}^{(2)}n_{\rm H II}+\alpha_{\rm He}^{(2)}n_{\rm He II})n_e,
\label{xi_rec}
\ee 
where $n_e$ is the number density of electrons, $n_{\rm H II}$ and 
$n_{\rm He II}$ are the number densities of H II and He II, respectively, while
$\alpha_{\rm H}^{(2)}$ and $\alpha_{\rm He}^{(2)}$ are the Case B
recombination coefficients for H II and He II, respectively, $N_{\rm a}$ is
the total number of atoms initially inside the minihalo, and $N_{\rm ion}$ is
the number of these atoms initially inside the minihalo which are ionized by
the evaporation process.  At time $t=t_{\rm ev}$, practically all atoms
originally in the minihalo are ionized, thus $N_{\rm ion}=N_a$ and
$\xi=\xi(t_{\rm ev})$ is the total number of ionized photons consumed per
minihalo atom. We have neglected the recombinations of He~III to He~II because
these generally contribute diffuse flux which is absorbed on the spot by H and
He.  

\citet{HAM01} estimated the evaporation times and photon consumption rates for
evaporating minihaloes analytically as follows.  The evaporation time $t_{\rm
  ev}$ was approximated by the sound-crossing time for the characteristic size
of the minihalo at the sound speed of the ionized gas, $t_{\rm
  sc}=2r_t/c_s(10^4\rm K)$, roughly consistent with our earlier results for
the simulation of the photoevaporation of a $10^7M_\odot$ minihalo
\citep{SRM97,SRM98,SR00a,SR00b,SR01,S01}. Haloes were assumed to be
dark-matter dominated, nonsingular isothermal spheres according to the
Truncated Isothermal Sphere (TIS) model \citep{SIR99,IS01}. For a minihalo of
mass $M$ at high redshift $z$, the TIS halo solution yields \be t_{\rm
  sc}=98\,{\rm Myr}\left({M_{7}}\right)^{1/3} \left(\frac{\Omega_0
    h^2}{0.15}\right)^{-1/3} \left({1+z}\right)_{10}^{-1}T_4^{-1/2},
\label{t_sc}
\ee using the adiabatic sound speed
$c_s(T)=11.7(T_4/\mu)^{1/2}=15.2\,T_4^{1/2}\rm km\,s^{-1}$, where $\mu=0.59$
is the mean molecular weight for fully ionized gas of H and He if the
abundance of He is $A(\rm He)=0.08$ by number relative to H. Based on this
evaporation time estimate, a na\"\i ve estimate of $\xi$ is then made possible
by further assuming that the minihalo has zero optical depth to
ionizing photons and is fully ionized instantaneously but remains static at 
its initial density for a time $t_{\rm sc}$ (the optically-thin, static 
approximation, hereafter referred to as ``OTS''). Ignoring the contribution 
of He to recombinations, equation~(\ref{xi_rec}) then yields 
\be 
\xi_{\rm OTS}=1+f\frac{C_{\rm
    int}\langle n_H\rangle\alpha_H^{(2)}}{1+\delta_{\rm TIS}}t_{\rm sc}
\label{xi_ham}
\ee 
\citep{HAM01} (where we have added the first term on the r.h.s. above to their
equation to account explicitly for the fact that each atom must first be
ionized once before it can recombine and be ionized again), where 
$\langle n_H\rangle$ is the mean H atom number
density inside a halo, $C_{\rm int}\equiv \langle n_{\rm H}^2\rangle/\langle
n_H\rangle^2=444^2$ is the effective clumping factor for the TIS, and
$1+\delta_{\rm TIS}=130.6$ is the average overdensity of a TIS halo with
respect to the cosmic mean background density.  The quantity $f$ is an
``efficiency factor'' introduced to take account of details neglected by the
OTS assumptions. Together, equations~(\ref{t_sc}) and (\ref{xi_ham}) yield 
\be
\xi_{\rm OTS}=1+206 fT_4^{-3/4}M_7^{1/3}\left(\frac{\Omega_0
    h^2}{0.15}\right)^{-1/3} \left({1+z}\right)_{10}^{2}.
\label{xi_numeric}
\ee 
Here, we have defined $M_7\equiv M/10^7M_\odot$ and $(1+z)_{10}\equiv
(1+z)/10$.  \cite{HAM01} calibrated the factor $f$ against an optically-thin
simulation of the expansion of a uniformly photoheated halo (i.e. no radiative
transfer), obtaining $f\approx1$. According to this OTS approximation,
$\xi_{\rm OTS}\gg1$ for minihaloes during cosmic reionization, thus the
presence of minihaloes during cosmic reionization would have dramatically
increased the number of ionizing photons needed to complete reionization
\citep{HAM01}. According to equation~(\ref{xi_numeric}), that is, each atom
which had collapsed into a minihalo before reionization ended would have
consumed many ionizing photons during its photoevaporation, a
disproportionately large share of the ionizing photon background compared to
that which would be available to ionize the uncollapsed atoms in the diffuse
IGM.

As shown by the detailed simulations in Paper I, for haloes of mass
$10^7M_\odot$ exposed  at $z=9$ to the typical level of ionizing radiation
expected during reionization, while the OTS
approximation yields a reasonable, rough estimate of $t_{\rm ev}$, it grossly
overestimates $\xi$, by factors of 30-50, depending on the ionizing source
spectrum. This more accurate determination of $\xi$ in Paper I by numerical
hydrodynamical simulations with radiative transfer nevertheless showed that
minihalo photoevaporation could still have consumed enough photons to increase
substantially the total number required per baryon to complete reionization.
By contrast, earlier estimates of the requirements for reionization of the IGM
had concluded that only about one photon per atom was required, but these
estimates were based on calculations without sufficient resolution to account
for the minihaloes \citep{G00a,MHR00}. It is therefore crucial to extend the
accurate determination of $\xi$ in Paper I to the full range of minihalo
masses, ionizing flux levels and redshifts expected during reionization, in
order to evaluate the global effect of minihaloes on the process of
reionization.

\section{The Calculation}
\label{calc_sect}

\subsection{Numerical Method and Initial Conditions}

We have performed a large number of simulations of the photoevaporation of
individual minihaloes which occurs when a weak, R-type intergalactic I-front
encounters a minihalo during cosmic reionization.  Our simulation method and
initial conditions were described in some detail in Paper I, so we only
briefly summarize these here.  Each halo is modelled as a nonsingular TIS 
\citep{IS01}, surrounded by self-similar
cosmological infall according to the solution of \citet{B85}.  We generalized
this self-similar infall solution to apply also to a low-density $\Lambda$CDM
background universe at high redshift \citep{IS01}.  At high-$z$, the halo
radius for the current background cosmology adopted here is
$r_t=0.754\,M_7^{1/3}(1+z_{\rm coll})_{10}^{-1}$ kpc, the total atomic number
density at the center is $n_0=3.2(1+z_{\rm coll})_{10}^3\,\rm cm^{-3}$, and
the halo virial temperature and velocity dispersion are $T_{\rm
  vir}=4000\,M_7^{2/3}(1+z_{\rm coll})_{10}$ K and $\sigma_V=
5.2\,M_7^{1/3}(1+z_{\rm coll})_{10}^{1/2}\,\rm km\,s^{-1}$, respectively.  For
haloes with $T_{\rm vir}\leq 10^4$K (minihaloes), the gas atoms are neutral. We
consider haloes in the mass range between the Jeans mass of the uncollapsed IGM
prior to reionization,
$M_J=5.7\times10^3\left({\Omega_0h^2}/{0.15}\right)^{-1/2}
\times\left({\Omega_bh^2}/{0.02}\right)^{-3/5}
\left[{(1+z)}_{10}\right]^{3/2}M_\odot$, and the mass for which $T_{\rm vir}=
10^4$K according to the TIS model, $M(10^4\rm
K)=3.95\times10^7(\Omega_0h^2/0.15)^{-1/2}[(1+z)_{10}]^{-3/2}$.

Our simulations use the two-dimensional, axisymmetric Eulerian gas-dynamics
code CORAL, with radiative-transfer, Adaptive Mesh Refinement (AMR), and the
van Leer Flux-Vector Splitting Algorithm, which resolves shocks well and
properly tracks I-fronts ranging from fast, supersonic R-type to sub-critical,
subsonic D-type, as described in detail in Paper I (and references therein).  
All our simulations were
set up as we described in detail in Paper I.  We fixed the size of the
computational box in all simulations presented here to be the same in units of
the halo radius (with the long side $\sim 10$ times larger than the halo
radius), $x_{\rm box}\propto r_{\rm vir}\propto M_{\rm halo}^{1/3}(1+z_{\rm
  coll})^{-1}$, so that the fully-refined grid resolution is equivalent in all
cases.  As in Paper I, we consider three different cases for the spectrum of
the external source of ionizing radiation: (1) starlight with a 50,000K
black-body spectrum (hereafter, referred to as ``BB5e4''), representative of
massive Population II stars; (2) starlight with a 100,000K black-body spectrum
(hereafter, ``BB1e5''), as expected for massive Population III stars; and (3)
QSO-like, with a power-law spectrum with spectral index $-1.8$ (hereafter,
``QSO''). As in Paper I, we express the unattenuated external flux of
ionizing photons as a dimensionless quantity, $F_0$, the flux in units of that
from a source emitting $N_{\rm ph}=10^{56}\rm s^{-1}$ ionizing photons per
second at a proper distance $d$ of 1 Mpc, or equivalently, emitting $10^{52}\,
\rm s^{-1}$ at a distance of 10 Kpc, so
$F_0\equiv F/\{10^{56}\,s^{-1}/[4\pi (1\, {\rm Mpc})^2]\}=N_{\rm ph}/d^2_{\rm
  Mpc}$. 
 
\subsection{Numerical Resolution}
To begin, we have studied the numerical convergence of our simulations with
increasing spatial grid resolution, for the quantities $t_{\rm ev}$ and $\xi$,
in order to establish the limits of robustness and reliability of our results.
We have performed a series of four simulations, with fixed halo parameters and
external flux level, with our fully-refined grid resolution $N_{\rm
  cells}=N_r\times N_x$, in 2D, axisymmetric $(r,x)$-coordinates, increasing
by factors of 2 from $128\times256$ up to $1024\times2048$, for each of the
three incident spectra.  The results are shown in Table~\ref{table_resol}.  We
note that with increasing resolution both the evaporation time $t_{\rm ev}$
and the number $\xi$ of ionizing photons consumed per atom decrease, leading
ultimately to convergence in both quantities at our highest resolution.  The
convergence is more readily achieved for the cases with harder spectra (QSO
and BB1e5) than for the softer spectrum case BB5e4.  In all cases $\xi$
converges faster than $t_{\rm ev}$.
\begin{table*}
\caption{Numerical convergence test for the evaporation time $t_{\rm ev}$ and
  photon  consumption rate per atom $\xi$ as functions of grid resolution for 
  minihalo of initial mass $10^7M_\odot$, exposed to initial flux of $F_0=1$, 
  starting at $z=9$.}
\begin{minipage}{3in}
  \centering
\begin{tabular}{@{}|c|c|c|c|c|c|}\hline
spectrum&$N_{\rm cell}$&$(\Delta x)_{\rm min} [pc]$&$r_t/(\Delta x)_{\rm min}$&$t_{\rm ev}$ [Myr]&$\xi$\\
\hline
BB5e4      & $128\times256$   &27.0 &23.4 & 465              &6.41 \\
BB5e4      & $256\times512 $  &13.5 &56.7 & 300              &5.73 \\
BB5e4      & $512\times1024$  &6.7  &113  & 195              &5.39 \\ 
BB5e4      & $1024\times2048$ &3.4  &227  & 150              &5.14\\
QSO        & $128\times256$   &27.0 &23.4 & 112              &6.20\\
QSO        & $256\times512$   &13.5 &56.7 & 103              &5.45 \\
QSO        & $512\times1024$  &6.7  &113  & 100              &5.20 \\ 
QSO        & $1024\times2048$ &3.4  &227  & 97               &4.95 \\ 
BB1e5      & $128\times256$   &27.0 &23.4 & 215              &5.16 \\
BB1e5      & $256\times512 $  &13.5 &56.7 & 178              &4.03 \\
BB1e5      & $512\times1024$  &6.7  &113  & 145              &3.39\\ 
BB1e5      & $1024\times2048$ &3.4  &227  & 128              &3.33\\ 
\hline
\end{tabular}
\label{table_resol}
\end{minipage}
\end{table*}

\subsection{Parameter space}
In order to study how the photoevaporation of minihaloes by global I-fronts
during reionization depends upon the halo mass, the external ionizing flux 
level and spectrum, and the redshift at which the I-front encounters the
minihalo, we have performed the following
large set of simulations. We considered a range of halo masses from a low of
$10^4M_\odot$, which is close to the Jeans mass at the corresponding epoch, to
$4\times10^7M_\odot$, close to $M(10^4\,K)$, the mass at which the halo virial
temperature is $T_{\rm vir}\sim 10^4$ K (e.g. $M(10^4\,K)=4\times10^7M_\odot$
at z=9). The intergalactic I-front which encountered the minihalo in each case
was assumed to reach the boundary of our simulation volume at a starting
redshift $z_i$ (and we assumed $z_{\rm coll}=z_i$ in defining the initial
parameters of the TIS model for the halo). We took $z_i$ to range from
$1+z_i=20$, consistent with the early start for reionization implied by the
recent WMAP results, to $1+z_i=7$, at which epoch the data from QSO absorption
line spectra suggest that reionization was just ending. The flux levels
assumed ranged from $F_0=0.01$ to $F_0=10^3$. We argued in Paper I that the
interval $F_0=0.01-100$ corresponds approximately to the typical range of
fluxes expected when a global I-front encountered minihaloes during the
reionization epoch. Here we extend this range up to $F_0=10^3$, to accommodate
cases of very high flux due either to a rare, very strong source or to the
minihalo's proximity to the source.  In all our simulations the ratio of box
size to halo radius was kept fixed at $L_{\rm box}/r_t\sim10$, which proved
sufficient to follow the evaporation flow until full evaporation was achieved.
Such a box size guaranteed that our transmissive boundary conditions were
self-consistent at all times, since the simulation volume (fixed over time in
proper coordinates) was always large enough to contain the sphere of radius
equal to the turn-around radius in the cosmological infall profile centered on
the minihalo as that radius increased over time.  We show the complete set of
simulation parameters and results for $t_{\rm ev}$ and $\xi$ in
Figure~\ref{scalings_fig}.  In order to make such a large set of simulations
feasible in terms of both computing time and storage requirements, we limited
the degree of refinement of the mesh so that the fully-refined mesh had
$512\times1024$ cells in all cases. Our numerical convergence tests above
showed that, at this resolution, the simulations are already largely
converged.  To correct for any remaining small differences from the
fully-resolved limit, we used the highest-resolution simulations in
Table~\ref{table_resol} to re-normalize our plotted results for $t_{\rm
  ev}$\footnote{For BB5e4 and BB1e5 cases in Figure~\ref{scalings_fig} which
  are not listed in Tables~\ref{table_resol} and \ref{table_checks}, the
  plotted points for $t_{\rm ev}$ are the actual values simulated at
  fully-refined grid resolution $512\times1024$, adjusted slightly to
  correspond to the converged limit of our results for the highest grid
  resolution of $1024\times2048$, based upon the convergence test results in
  Table~\ref{table_resol}. The adjusted values are 0.77 (0.88) times the
  simulation values for resolution $512\times1024$ for cases BB5e4 (BB1e5), 
  respectively.}.  
Furthermore, in order to verify this approach, we performed a limited set 
of additional higher-resolution ($1024\times2048$ fully refined)
simulations for different halo masses and external fluxes as presented in
Table~\ref{table_checks}.
\begin{table*}
\caption{Evaporation times and ionizing photon consumption rates per atom for 
 several simulations at highest grid resolution ($1024\times2048$ fully
 refined), all for $z_i=9$ and $r_t/(\Delta x)_{\rm  min}=227$. }
\begin{minipage}{3in}
  \centering
\begin{tabular}{@{}|l|c|c|c|c|c|c|c|}\hline
spectrum&$F_0$ &$M_7$   &$(\Delta x)_{\rm min} [\rm pc]$&$t_{\rm ev}$ [Myr]&$t_{\rm ev}$ [Myr] (fit)& $\xi$& $\xi$ (fit)\\
\hline 
BB5e4 & 1        &$10^{-3}$&0.34& 8.5  & 7.5       & 1.75           & 1.71\\
BB5e4 & 10       &4        &5.4 & 170  & 137       & 10.6           & 11.2\\
BB5e4 & 0.5      &1        &3.4 & 210  & 193       & 4.7            & 4.8 \\
BB5e4 & 10       &1        &3.4 & 85   & 75        & 6.85           & 7.3 \\
BB5e4 & $10^2$   &1        &3.4 & 49   & 47        & 9.07           & 8.5 \\
BB5e4 & $10^3$   &1        &3.4 & 37   & 38        & 9.80           & 8.3 \\
QSO    & $10^2$   &1        &3.4 & 25   & 21        & 10.9           & 11.0\\
QSO    & $10^3$   &1        &3.4 & 9    & 10        & 13.6           & 14.6\\
BB1e5 & $10^2$   &1        &3.4 & 41   & 41        & 5.28           & 5.7 \\
BB1e5 & $10^3$   &1        &3.4 & 29   & 34        & 6.41           & 6.2 \\
\hline
\end{tabular}
\label{table_checks}
\end{minipage}
\end{table*}

\section{Results}
\label{results_sect}

\subsection{Evaporation time}
\label{tev_sect}
\begin{figure}
  \includegraphics[width=3.5in]{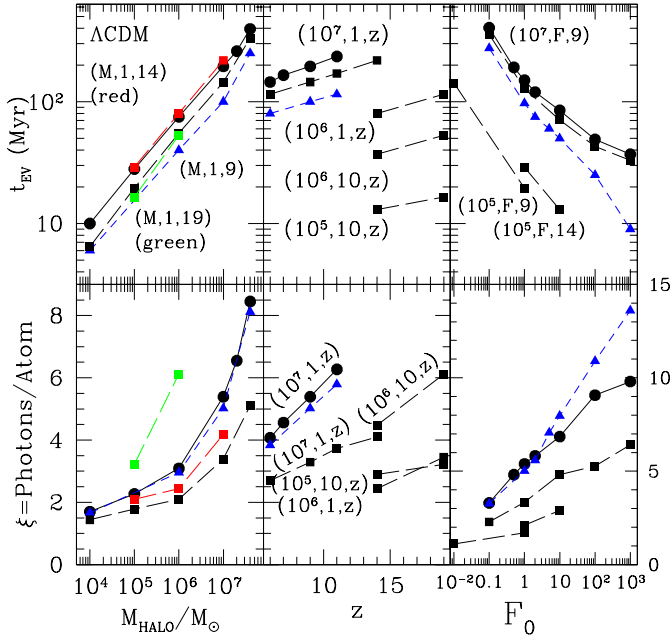}
\caption{Photoevaporation times $t_{\rm ev}$ for individual minihaloes 
  (top panels) and total ionizing photon consumption per halo atom $\xi$ 
  (bottom panels) vs. halo mass $M$ (left), redshift $z$ at which I-front 
  first encounters the minihalo (middle) and dimensionless ionizing flux 
  $F_0$ (right), for the three different types of source spectra, BB5e4 
  (circles), QSO (triangles) and BB1e5 (squares), labelled by $(M,F_0,z)$ 
  to indicate the parameters which we kept fixed in each case as we varied 
  the quantity on the x-axis. The red (green) lines on the two left panels 
  correspond to redshift $1+z=15$ ($1+z=20$). Also, note the different 
  vertical scale of the lower right panel only.}
\label{scalings_fig}
\end{figure}

\begin{figure}
  \includegraphics[width=3.5in]{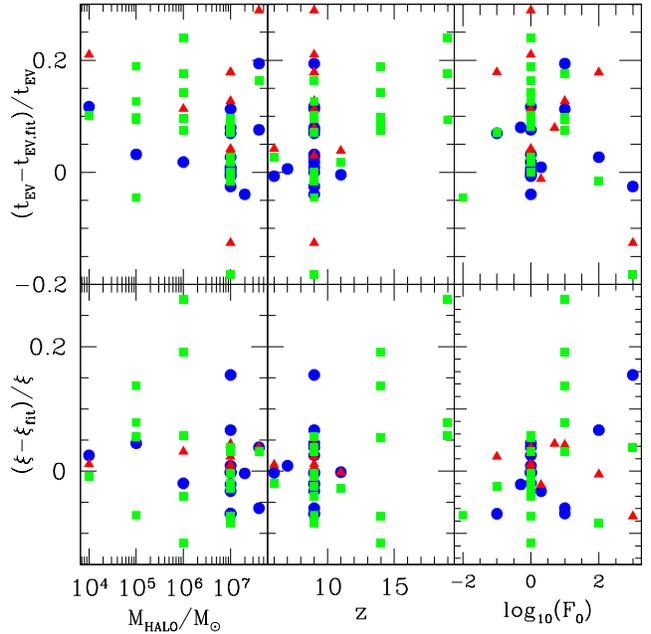}
\caption{Fractional errors of the fitting formulae in 
  equations~(\ref{tev_fit}) and (\ref{xi_fit}) compared with the simulation
  results for each of the cases plotted in Figure~\ref{scalings_fig}, for the
  evaporation times (top panels) and total ionizing photon consumption
  rates $\xi$ (bottom panels), respectively, vs. halo mass $M$ (left), 
  redshift $z$
  at which I-front encounters the minihalo (middle) and dimensionless ionizing
  flux $F_0$ (right), for BB5e4 (blue circles), QSO (red triangles) and 
  BB1e5 (green squares) spectra.}
\label{errors_fig}
\end{figure}

\begin{figure}
  \includegraphics[width=3.5in]{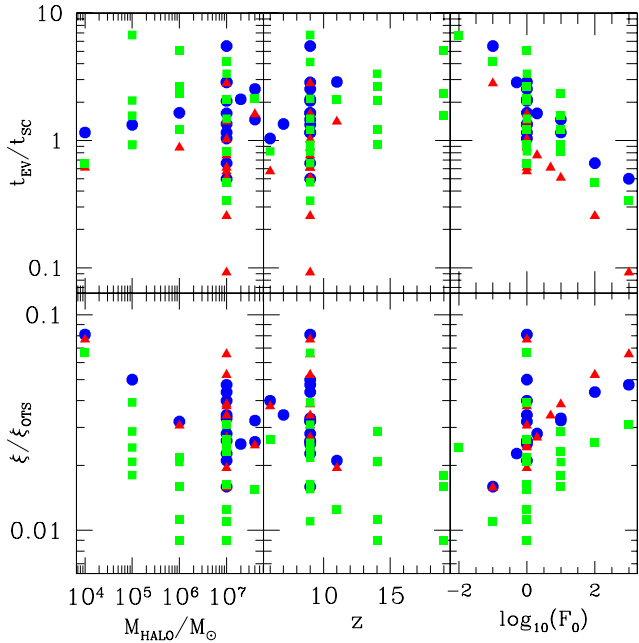}
\caption{Comparison of simulation results for the evaporation times and photon
  consumption rates with predictions of the OTS approximation. Ratios of
  $t_{\rm ev}/t_{\rm sc}$ (top panels) and $\xi/\xi_{\rm OTS}$ (bottom panels)
  vs. halo mass $M$ (left), redshift $z$ at which I-front encounters the
  minihalo (middle) and dimensionless ionizing flux $F_0$ (right), for each 
  of the
  cases plotted in Figure~\ref{scalings_fig} for BB5e4 (blue circles), QSO
  (red triangles) and BB1e5 (green squares) spectra.}
\label{OTS_fig}
\end{figure}

Our simulation results for the evaporation time $t_{\rm ev}$ are plotted for
all cases in Figure~\ref{scalings_fig} (upper panels).  The evaporation time
increases significantly with increasing halo mass. For the smallest minihaloes
$t_{\rm ev}\sim10$ Myr, much shorter than the Hubble time at that epoch, while
for the larger minihaloes $t_{\rm ev}\sim 100$ Myr, closer to, but still
shorter than the current Hubble time. The redshift dependence of the
evaporation time is even stronger than the mass dependence, with $t_{\rm ev}$
increasing with redshift. However, due to the relatively small range of the
relevant redshifts, the overall variation of $t_{\rm ev}$ with redshift is
only by factor of $\sim3$ for fixed halo mass and external flux. Since the
Hubble time $t_H\propto (1+z)^{-1.5}$, the evaporation times of
higher-redshift haloes are a much larger fraction of the Hubble time, although
still generally smaller than $t_H$. Finally, the evaporation times depend
inversely on the level of the external ionizing flux, becoming extremely long
($t_{\rm ev}\sim$ few hundreds of Myr) for low flux levels ($F_0\simless
0.1$), and quite short ($t_{\rm ev}\simless30-40$ Myr) for high flux levels
($F_0\simgreat10^2$).

The logarithmic slope of the dependence of $t_{\rm ev}$ on mass and the linear
slope of its dependence on the initial redshift show no significant
dependences on the source spectrum for the range of masses, fluxes and
redshifts we have studied, although the overall normalization does vary. The
dependence of $t_{\rm ev}$ on the flux is more complicated, with a slope which
varies with $F_0$.

A simple analytical fit to these results is given by 
\be 
t_{\rm ev}=A M_7^{B}
F_0^{C+D \log_{10} F_0}\left[E+F\left(\frac{1+z}{10}\right)\right]\,{\rm Myr}
\label{tev_fit}
\ee 
where $A=(150,97,128)$, $B=(0.434,0.437,0.465)$,
$C=(-0.35,-0.357,-0.358)$, $D=(0.05,0.01,0.056)$, $E=(0.1,0.3,0.24)$, and
$F=(0.9,0.7,0.76)$ for cases (BB5e4, QSO, BB1e5), respectively.  We have fit
the functional dependences based on all the points plotted in
Figure~\ref{scalings_fig}.  The relative errors of these fitting formulae are
plotted in Figure~\ref{errors_fig} (upper panels) for all the cases in
Figure~\ref{scalings_fig}. Table~\ref{table_checks} shows the comparison
between the simulation values and the fitting formulae for several of our
highest-resolution simulations. In most cases the errors are $\simless10\%$,
and in all but two cases the errors are $\simless20\%$.

We have compared our simulation results for $t_{\rm ev}$ with the OTS estimate
of this quantity by \citet{HAM01} in Figure~\ref{OTS_fig}. The OTS
approximation, recall, assumes that the sound-crossing time of the halo once
it is photoheated by ionizing radiation, $t_{\rm sc}=2r_t/c_s$, where $c_s$ is
the speed of sound at $10^4$ K, is a good approximation to the evaporation
time. According to equation~(\ref{t_sc}), this means $t_{\rm ev,OTS}=t_{\rm
  sc}\propto r_t\propto M^{1/3}(1+z)^{-1}$, so the OTS evaporation time for a
fixed halo mass {\it decreases} with increasing redshift in inverse proportion
to the redshift, rather than {\it increasing} linearly with redshift, as in
our numerical results, according to the scaling in equation~(\ref{tev_fit}), 
while the increase of
$t_{\rm ev}$ with halo mass in the OTS prediction is somewhat less steep 
than our
result. Finally, the OTS approximation completely ignores the significant
dependence of $t_{\rm ev}$ on the magnitude and spectrum of the ionizing flux,
making $t_{\rm sc}$ a significant underestimate of $t_{\rm ev}$ (by up to an
order of magnitude) for low fluxes and a similarly significant overestimate
for high fluxes.
 
\subsection{Ionizing photon consumption}
\label{xi_sect}
The simulation results for the number of ionizing photons per minihalo atom
required to evaporate a minihalo, $\xi$, are plotted in
Figure~\ref{scalings_fig} (lower set of panels). The photon consumption per
atom increases steeply with increasing halo mass. For the smallest minihaloes,
$\xi\sim2$, regardless of the spectrum of the ionizing source, i.e. each atom
on average recombines just once during the evaporation of these haloes, or,
equivalently, is ionized twice by the time it is expelled from the halo in the
evaporative wind. On the other hand, for the larger minihaloes, the photon
consumption rate is significantly larger, $\xi\sim 5-8$, so each atom in these
haloes recombines multiple times on average during the evaporation of the
halo.  A BB1e5 source evaporates a large minihalo a factor of $\sim 2$ more
efficiently in terms of photon consumption than a QSO or BB5e4 source,
although the differences in efficiency between different spectra almost
disappear for the smaller minihaloes.  The ionizing photon consumption rate
grows approximately linearly with redshift, and hence is noticeably higher at
higher redshifts. Finally, opposite to the trend for the evaporation time, the
photon consumption rate {\it grows} with increasing flux, with only 2-3
photons per atom needed in the case of low ionizing flux ($F_0\simless 0.1$),
but up to 7-16 needed in the case of high flux levels ($F_0\simgreat10^2$).
Therefore, and somewhat counter-intuitively, the low flux levels are much more
efficient than the high flux levels, although the evaporation process takes
much longer in the former than in the latter cases.

A good fit to these results is given by \be \xi=1+A M_7^{B+C\log_{10} M_7}
F_0^{D+E\log_{10} F_0}\left[F+G\left(\frac{1+z}{10}\right)\right],
\label{xi_fit}
\ee 
where $A=(4.4,4.0,2.4)$, $B=(0.334,0.364,0.338)$, $C=(0.023,0.033,0.032)$,\\
$D=(0.199, 0.24, 0.219)$, $E = (-0.042,-0.021,-0.036)$, $F=(0,0,0.1)$,
$G=(1,1,0.9)$ for the BB5e4, QSO, and BB1e5 spectra, respectively. The
relative errors of these scaling laws are plotted in Figure~\ref{errors_fig},
lower panels. In most cases the errors are $\simless10\%$, and in all cases
but one the errors are below $20\%$.

We have compared our simulation results for ionizing photon consumption with
the predictions of the OTS approximation in Figure~\ref{OTS_fig}.  The OTS
prediction for $\xi$ is given by equation~(\ref{xi_numeric}) with $f=1$. The
number of recombinations per atom, $\xi_{\rm OTS}-1$, scales with mass and
redshift as $M^{1/3}(1+z)^2$. While the logarithmic slope of this OTS
dependence on the halo mass is roughly similar to that of our simulations
according to equation~(\ref{xi_fit}), the increase of $\xi$ with redshift 
of the OTS prediction is much stronger than that of the simulation results. In
addition, the dependence of $\xi$ on the ionizing flux and spectrum is quite
strong in the numerical results, while it is completely ignored in the OTS
approximation. Unlike the evaporation time, $t_{\rm ev}$, however, the overall
normalization of $\xi$ in the OTS approximation is overestimated by at least
an order of magnitude in all cases, as compared with the simulation results,
and by more than two orders of magnitude in some cases, particularly at high
redshift and/or low ionizing flux levels.

\subsection{Evolution of the neutral mass fraction}
\label{Mn_sect}
\begin{figure}
  \includegraphics[width=3.5in]{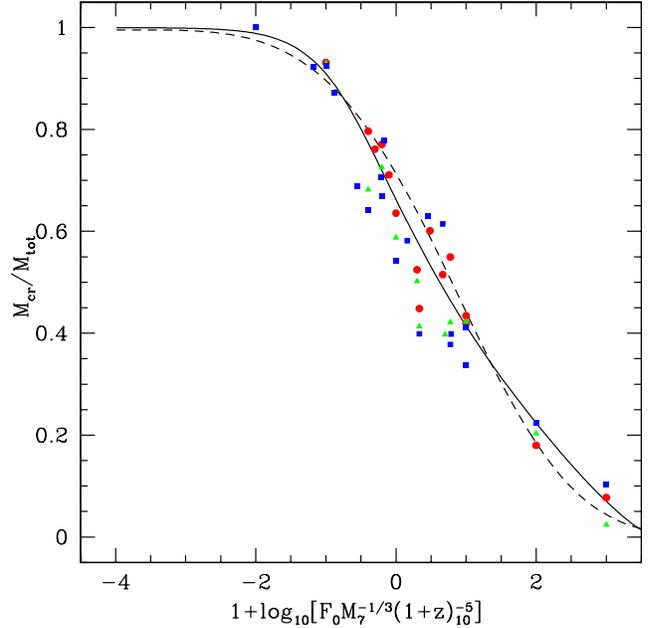}
  \caption{Neutral mass fraction of minihalo at the moment of I-front
    trapping, $M_{\rm cr}$. Points show results from simulations with BB5e4
  (circles), QSO (triangles) and BB1e5 (squares) spectra, respectively. For
  comparison, we plot the analytical prediction for this quantity derived from
  the Inverse Str\"omgren Sphere (ISL) approximation in Paper I (with argument
  on x-axis shifted as described in text), in which 
  $M_{\rm cr,analyt}\equiv1-M_{\rm ISL}/M_{\rm tot}$ (solid line), and a
  convenient fit to the ISL approximation  and simulation results, according
  to equation~(\ref{mcr_fit}) (dashed line).}
\label{m_cr_fit}
\end{figure}
As described in Paper I, the first phase of the encounter between an
intergalactic I-front and a minihalo is the weak, R-type phase, during which
the I-front enters the halo and photoionizes an outer layer of halo gas,
predominantly on the side facing the ionizing source, before the gas has time
to respond hydrodynamically and move. This phase ends when the I-front slows
to the R-critical speed and makes a transition to D-type, thereby trapping the
I-front until it can burn through the remaining neutral, shielded minihalo gas
and evaporate it. We identify the fraction of the original minihalo gas mass
which was photoionized in this initial R-type phase when the front became
R-critical as $M_{\rm cr}$. We have plotted in Figure~\ref{m_cr_fit} the
values of $M_{\rm cr}$ for our simulation cases in Figure~\ref{scalings_fig}.
In Paper I, we derived an analytical approximation which can be used to
estimate $M_{\rm cr}$, called the Inverse Str\"omgren Layer (ISL)
approximation, where ISL is the ionized boundary layer of the minihalo defined
by the ``outside-in'' Str\"omgren length for each impact parameter (see Paper
I for details). In the static, ionization-equilibrium approximation, the mass 
contained
in this boundary layer, $M_{\rm ISL}$, which is readily obtained for our TIS
model by numerical 
integration as shown in Paper I, can be used to estimate $M_{\rm cr}$ as the 
complement of $M_{\rm ISL}$, $M_{\rm cr}=M_{\rm tot}-M_{\rm ISL}$.

In Paper I, we compared our simulation results for the ionized mass fraction
at the transition from R-type to D-type for a $10^7M_\odot$ halo exposed at
$z_i=9$ to a source of flux $F_0=1$ and found an excellent agreement with 
$M_{\rm ISL}$. Here we extend the comparison to determine how well that 
agreement
holds up when we vary the halo parameters, source flux level, and redshift. We
find that the dependence of the simulation results on these parameters follows
that predicted by the ISL approximation fairly well, where the latter predicts
that $M_{\rm ISL}$ should depend upon $(M,F_0,z)$ only in the combination
$F_0M^{-1/3}(1+z)^{-5}$. We find, however, that the full range of simulation
results for $M_{\rm cr}$ is better fit by this ISL scaling law if we multiply
the quantity $F_0M^{-1/3}(1+z)^{-5}$ in the original ISL approximation by 10.
The excellent agreement between this modified ISL scaling law for $M_{\rm cr}$
and the simulation results is shown in Figure~\ref{m_cr_fit}. We find that the
following analytical expression provides a reasonable fit both to the ISL
curve (with shifted argument) and the simulation results, \be M_{\rm
  cr,approx}=0.9954\exp(-0.0013x^4),
\label{mcr_fit}
\ee where $x\equiv 5+\log_{10}[F_0M_7^{-1/3}(1+z)_{10}^{-5}]$.

\begin{figure}
  \includegraphics[width=3.5in]{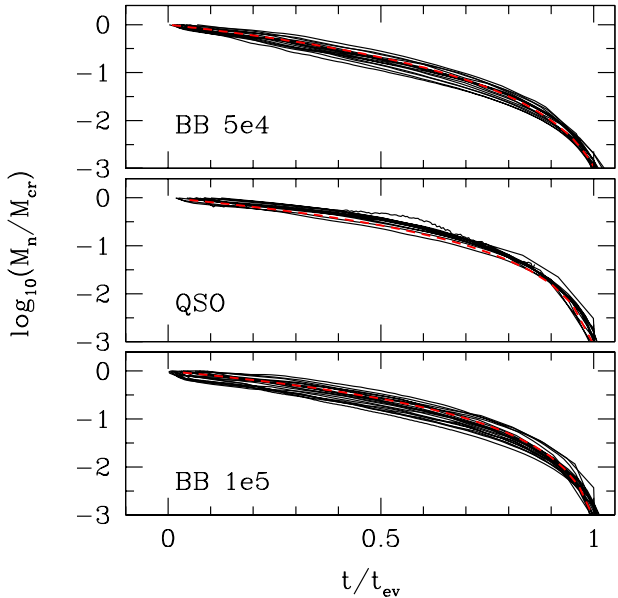}
\caption{Evolution of the neutral mass fraction $M_n/M_{\rm cr}$
  from the end of the initial R-type phase of the I-front and its 
  conversion from R-type
  to D-type to the final evaporation of the halo, for simulations with BB5e4
  (top), QSO (middle) and BB1e5 (bottom) spectra (solid lines). We also show
  the best fits in each case, as described in the text (long-dashed red
  lines).}
\label{mass_evol}
\end{figure}
In Fig.~\ref{mass_evol} we show the evolution of the neutral mass fraction,
$M_n/M_{\rm cr}$ vs. $t/t_{\rm ev}$ as the minihalo evaporates, for all of the
cases plotted in Figure~\ref{scalings_fig}. This time-variation of the neutral 
mass fraction of the evaporating minihaloes is apparently a nearly universal
function, when the neutral mass $M_n(t)$ is expressed in units of $M_{\rm cr}$
and $t$ is expressed in units of $t_{\rm ev}$ for each case, independent of
the halo parameters and external flux, but allows for some dependence on the
spectrum of the external source. The scatter in all cases is largely due to
the decrease of the flux level in time due to cosmological expansion, for a
source located at fixed comoving distance from the target minihalo.
Accordingly, the relation is tightest in the QSO case, since in these runs the
evaporation times are shorter. The average shape of the neutral mass
time-dependence function is well-fit by 
\be 
M_n(t)=M_{\rm cr}\left(1-\frac{t}{At_{\rm ev}}\right)^{B},
\label{mass_loss_numeric}
\ee 
where $A=(1.07,1.03,1.03)$ and $B=(2.5,2,2)$ for the BB5e4, QSO and BB1e5 
spectra, respectively (red dashed lines in Fig.~\ref{mass_evol}).

\section{Implications for Reionization and Conclusions}
\label{conclusions_sect}

The widespread presence of minihaloes filled with neutral gas during the epoch
of reionization made encounters between the weak, R-type intergalactic
I-fronts which ionized the IGM and these minihaloes a common occurrence. This
interaction affected both the minihaloes and the I-fronts profoundly. The
trapping of the I-fronts by minihaloes which cover the sky as seen by the more
massive haloes which are generally believed to have been the primary sources 
of cosmic
reionization slowed the advance of those I-fronts in the IGM, at large, and
wasted some of the ionizing radiation which would otherwise have been
available to ionize that IGM. The minihaloes, in turn, were transformed by the
photoevaporation which followed their trapping of the intergalactic I-fronts,
into barren, dark-matter haloes devoid of baryons. In order to explore this
process and its implications for cosmic reionization, we performed a large set
of high resolution gas dynamical simulations which include radiative transfer,
the first in their kind. In Paper I, we discussed our results in detail for
the illustrative case of a $10^7M_\odot$ halo which is overrun by an
intergalactic I-front at $z_i=9$, from a source of flux $F_0=1$, for different
source spectra. Here we have summarized a much larger set of simulations 
designed to quantify the dependence of minihalo photoevaporation on the halo 
mass, source flux level and spectrum, and redshift.
In Paper I, we discussed the comparison between our simulation results for
$10^7M_\odot$ haloes and previous related work. Here, we extend this
comparison to include the broader set of cases presented in this paper. 

We noted in Paper I that the approximate analysis of \citet{BL99} which was 
used to argue that photoevaporation of minihaloes affected a significant part 
of the baryon fraction collapsed into haloes before the end of reionization was
inconsistent with the results of our detailed numerical simulations for
$10^7M_\odot$ haloes. \citet{BL99} had modelled this process by a static
approximation like the ISL approximation in Paper I, to determine what portion
of the halo's mass is shielded from external ionizing photons and what portion
is exposed to photoheating. They then assumed that only that portion which
was instantaneously heated in this way, enough to unbind it from the halo
gravitational potential well, would evaporate from the halo. From this static
approximation, they concluded that only minihaloes smaller than 
$\sim {\rm few}\times10^5-10^6\,M_\odot$ (depending on the redshift and  
the source spectrum) were evaporated completely during reionization, while the
larger minihaloes would have retained a significant fraction of their gas (up
to $\sim 50\%$ of the gas for haloes of mass $\approx$few$\times10^7M_\odot$). 
In Paper
I we demonstrated that this model does not account properly for the dynamics
of photoevaporation, and, as a result, it fails to anticipate the fact that
even minihaloes as large in mass as $M=10^7M_\odot$ must ultimately evaporate 
completely. The larger set of simulations we have presented here show further 
that this neglect of dynamics by \citet{BL99} in estimating the dependence of
photoevaporation on halo mass is not correct; all minihaloes exposed to
ionizing radiation during reionization evaporate completely. All of the gas 
initially in a minihalo would eventually have become unbound, outflowing with 
speeds of $v_{\rm wind}=20-40\,{\rm km\, s^{-1}}$, leaving behind just a dark 
halo. The static approximation fails because it ignores
the dynamical nature of photoevaporation. As evaporation proceeds, layers
of gas are continuously stripped away, exposing the gas layers within to the
ionizing radiation. Thus, although the inner halo region self-shields
initially, all the gas is eventually ``unshielded'' and photoheated to 
$T>10^4$ K, well above the halo virial temperature, and, hence, boils out of 
the halo.

In Paper I we pointed out that simulations of the photoevaporation of
$10^7M_\odot$ haloes by an intergalactic I-front during reionization did not
support the suggestion by \citet{C01} that external ionizing radiation can 
cause the implosion of the minihalo gas, leading to globular cluster
formation. We did not observe such an effect in our simulations reported
there. The broad range of cases considered here allow us to extend this 
comparison to the full range of
parameter space expected for minihaloes exposed to the effects of
externally-driven I-fronts during reionization. In no case have we yet found
evidence of the implosion effect predicted by \citet{C01}.

Our simulations show that the number of ionizing photons per minihalo atom 
needed to photoevaporate the minihalo is typically in the range
$\xi\sim2-10$, significantly larger than the value of $\xi\sim1$ previously 
estimated to be sufficient to reionize the IGM 
\citep{G00a,MHR00}.  On the other hand, we have also shown that recent claims 
\citep{HAM01} that the photoevaporation of minihaloes can require up to
hundreds of ionizing photons per atom are very significantly overestimating 
$\xi$.
\begin{figure}
  \includegraphics[width=4.5in]{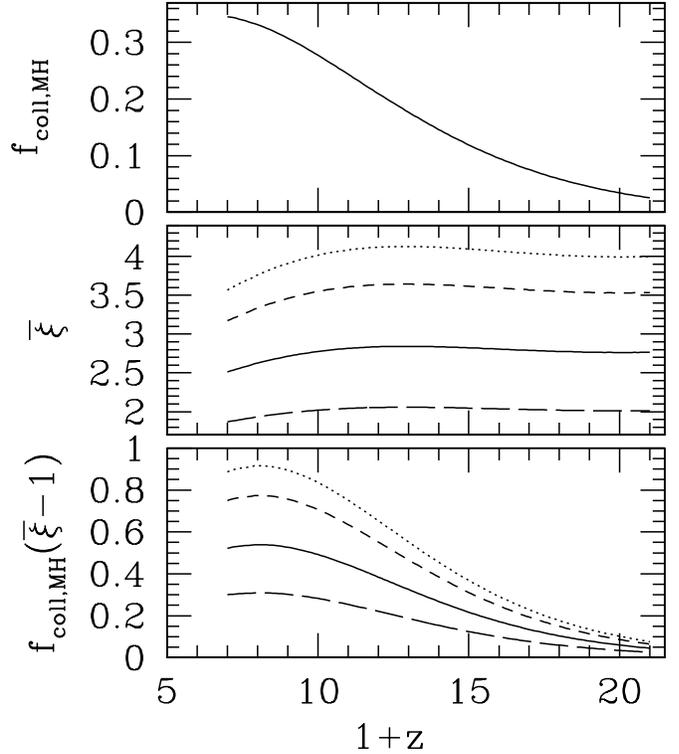}
\caption{Collapsed fraction in minihaloes $f_{\rm coll,MH}$ (top panel) for
  the $\Lambda$CDM universe, mean photon consumption per
  minihalo atom, $\bar{\xi}$, due to photoevaporation of minihaloes  
  (middle panel), and mean excess photon consumption per total atom 
  (i.e. all atoms, including both minihaloes and IGM) compared to the nominal
  requirement of one per atom (bottom panel).   Results are for
  ionizing fluxes $F_0=0.1$ (long-dashed), 1 (solid), 10 (short-dashed) and
  100 (dotted), for case BB5e4.}                       
\label{fcoll_xi}
\end{figure}

We have quantified here the dependence of evaporation times and photon 
consumption rates on the halo mass, source flux level and spectrum, and 
redshift of encounter between I-front and minihalo. We have provided 
convenient fitting formulae for our simulation results as functions of the
parameters $(M,F_0,z)$, which should be useful in further analyses
of the role of this process in cosmic reionization. To go beyond this step, it
is necessary to determine the distribution in time and space of these
parameters. The effect on reionization will depend upon the luminosity
function of source haloes and the mass function of minihaloes as they evolve in
time and fluctuate in space, including the clustering of both the sources and
minihaloes. These may depend, in turn, upon the feedback of reionization on 
galaxy formation, so it is likely to be necessary to solve the problem of
reionization in some detail before the true impact of minihaloes as screens
and photon sinks is known. That is well beyond the scope of this paper.
Here, instead, we shall provide a simple first estimate of the average effect 
of minihalo photon
consumption on the photon budget required for reionization, based on the mean
Press-Schechter (PS) distribution of minihaloes, as follows.

Let the ionizing photon consumption rate per minihalo atom be given for a
single minihalo by $\xi(M,z,F_0)$. Then its average over all minihaloes at
redshift $z$ is given by 
\be
\bar{\xi}(z,F_0)=\frac1{\rho_{tot}f_{\rm
    coll,MH}}\int_{M_{\rm min}}^{M_{\rm max}}\xi(M,z,F_0)M\frac{dn(M,z)}{dM}dM,
\label{PS_average}
\ee 
where ${dn(M,z)}/{dM}$ is the well-known PS mass function of haloes,
\be
f_{\rm coll,MH}\equiv\frac1{\rho_{tot}}\int_{M_{\rm min}}^{M_{\rm max}}
M\frac{dn(M,z)}{dM}dM
\ee
is the collapsed mass fraction in minihaloes, $\rho_{tot}$ is the mean matter 
density at the corresponding redshift, $M_{\rm min}$ is the Jeans mass in the
uncollapsed IGM, $M_J$, and
$M_{\rm max}=M(10^4\rm K)$, is the halo mass corresponding to virial
temperature $10^4$ K. Assuming that $\xi(M,z,F_0)$ is given
by the fitting formula in equation~(\ref{xi_fit}), we can calculate
$\bar{\xi}(z,F_0)$ for any given redshift $z$ and ionizing flux $F_0$. Results
for $F_0=0.1$, 1, 10 and 100 and $7\leq 1+z\leq21$ are shown in
Figure~\ref{fcoll_xi}, for case BB5e4. We see that $\bar{\xi}$ depends 
strongly on the ionizing
flux ($\bar\xi \propto F_0^{0.199-0.042\log_{10} F_0}$, in
equation~[\ref{xi_fit}]), but is approximately independent of the
redshift. Thus, the contribution of minihalo photoevaporation to the mean
global photon consumption rate is largely dictated by the ionizing flux 
levels and the current collapsed fraction in minihaloes and can require up to 
$\sim1$ additional ionizing photon per atom in the universe (both minihalo and
IGM atoms) to finish reionization, as compared with the requirement when
minihaloes are neglected, for the flux levels we
considered here. Hence, minihalo photoevaporation can potentially double the
number of ionizing photons required to reionize the universe. 

We defer the complete treatment of minihalo and source bias
which would account for the spatially and temporally-varying fluxes during
reionization to a future paper (Iliev, Scannapieco \& Shapiro, in
preparation). A brief summary of our first results along these lines can be
found in \citet{ISS04}, where it is shown that the statistical bias which
causes the minihalo mass function to be enhanced around source haloes relative
to the mean PS mass function boosts the relative contribution of minihaloes as
photon sinks and compensates for the declining mean collapsed fraction at high
redshift.

This shows that minihalo photoevaporation will be an important feature of
reionization, causing a modest, but significant slow down of the global
I-fronts and delaying their overlap. Only a more detailed, self-consistent
treatment of the global process for a given reionization scenario will
determine if reionization was photoevaporation-{\it dominated}, since the
influence of minihaloes is strongly dependent on both the flux and, through the
evolving collapsed fraction, the redshift at which the I-fronts encountered the
minihaloes. Additionally, minihalo formation in some places and at some epochs
might be partially suppressed by e.g. an early X-ray background, double
reionization, or reduced small scale power. Such different scenarios might be
distinguished observationally by detecting the fluctuations of the redshifted
21-cm emission or absorption lines from minihaloes \citep{ISFM02,ISMS03,C03} or
the kind of absorption line signatures at UV and optical wavelengths discussed 
in Paper I.

\section*{Acknowledgments}

We thank Andrea Ferrara and Garrelt Mellema for many useful discussions. This
research was supported by NSF grant INT-0003682 from the International
Research Fellowship Program and the Office of Multidisciplinary Activities of
the Directorate for Mathematical and Physical Sciences, the Research and
Training Network "The Physics of the Intergalactic Medium" set up by the
European Community under the contract HPRN-CT2000-00126 RG29185, NASA ATP 
grants NAG5-10825 and NNG04G177G and Texas Advanced Research Program 
grant 3658-0624-1999.

\end{document}